\documentstyle[prl,aps,floats,epsf,color]{revtex}
\begin{document}
\baselineskip=12pt
\def\be{\begin{equation}}
\def\ee{\end{equation}}
\def\bea{\begin{eqnarray}}
\def\eea{\end{eqnarray}}
\def\orc{\Omega_{r_c}}
\def\om{\Omega_{\text{m}}}
\def\E{{\rm e}}
\def\bearst{\begin{eqnarray*}}
\def\eearst{\end{eqnarray*}}
\def\peleven{\parbox{11cm}}
\def\peffec{\peight{\bearst\eearst}\hfill\peleven}
\def\pspace{\peight{\bearst\eearst}\hfill}
\def\ptwelve{\parbox{12cm}}
\def\peight{\parbox{8mm}}
\twocolumn[\hsize\textwidth\columnwidth\hsize\csname@twocolumnfalse\endcsname

\title{ Inverse problem: Reconstruction of modified gravity action in Palatini formalism by Supernova Type Ia data}

\author{ Shant Baghram, Sohrab Rahvar}
\address{Department of Physics, Sharif University of
Technology, P.O.Box 11365--9161, Tehran, Iran}

\vskip 1cm

 \maketitle
\begin{abstract}
%{\it{CONTEXT}}:
We introduce in $f(R)$ gravity--Palatini formalism the method of
inverse problem to extract the action from the expansion history
of the universe. First, we use an ansatz for the scale factor and
apply the inverse method to derive an appropriate action for the
gravity. In the second step we use the Supernova Type Ia data set
from the Union sample and obtain a smoothed function for the
Hubble parameter up to the redshift~1.7. We apply the smoothed
Hubble parameter in the inverse approach and reconstruct the
corresponding action in $f(R)$ gravity. In the next step we
investigate the viability of reconstruction method, doing a
Monte-Carlo simulation we generate synthetic SNIa data with the
quality of union sample and show that roughly more than 1500 SNIa
data is essential to reconstruct correct action. Finally with the
enough SNIa data, we propose two diagnosis in order to
distinguish between the $\Lambda$CDM model and an alternative
theory for the acceleration of the universe.

\end{abstract}

\newpage
]

\section{Introduction}
Combination of SNIa+CMB data shows that universe is in the
positive acceleration phase \cite{R04,Dun09}. This result is in
contradiction with our expectation from the behavior of the
ordinary matter. The simplest solution is assuming a cosmological
constant in the Einstein field equation as a constant of
integration \cite{Adam98}. While this simple modification
explains the observational data\cite{Davis07}, however the
cosmological constant suffers from the fine tuning and
coincidence problems\cite{Wienberg89}. One of the solutions is
introducing a scalar field which provides a time dependent
negative equation of state \cite{Peeb03}.

The other possibility is the modification of the gravity law in
such a way that it behaves as standard General Relativity in
strong gravitational regimes and repulse particles in the low
density cosmological scales \cite{Carroll04}. The modified
gravity models can be examined with three category of observations
of (a) cosmological dynamics \cite{Jain08}, (b) local gravity
\cite{Faraoni07} and (c) the evolution of large scale structure
\cite{Koiv07}.

In this work we use SNIa data to reconstruct an appropriate $f(R)$
gravity model in Palatini formalism \cite{Fay07} with the inverse
method. The method is the extension of the work introduced for
the metric formalism by Rahvar and Sobouti in \cite{rahvar}. The
inverse method also is introduced in the work by Capozziello et
al.{\cite{Cappo2005}} in the metric formalism. In this method we
need to know the dynamics of Hubble parameter from the
observational data. Many methods for the extraction of the Hubble
parameter have been introduced in the literature
\cite{Starobinsky98}. Here we use the smoothing method suggested
by Shafieloo et al.\cite{shafieloo} and apply it to the SNIa
Union sample \cite{Kowalski}. We use $H(z)$ in the inverse method
algorithm to reconstruct the corresponding action. We test the
reliability of this method by doing a Monte-Carlo simulation and
generating the synthetic SNIa data according to an action and
comparing the reconstructed action with the original one. Finally
we introduces two diagnosis for distinguishing standard
$\Lambda$CMD model from the alternative models.

The structure of this article is as follows: In section
\ref{section2} we introduce $f(R)$ modified gravity in Palatini
formalism, derive the equation of motion and obtain the dependence
of Hubble parameter to the Ricci scalar and scale factor. In
section \ref{section3} we introduce the method of inverse problem
in Palatini $f(R)$ gravity. In section \ref{section4} we use the
method to the real data of SNIa, smoothing the supernova data we
extract the Hubble parameter in terms of redshift and apply it to
extract the action \cite{shafieloo}. Also in order to show the
level of confidence of our results, we simulate 100 realization
of SNIa data to extract the Hubble parameter and compare it with
that obtained directly from fitting to the model. In the section
\ref{section5}, we examine the viability of reconstruction method
and dependence of the results to the number of SNIa data. In
section \ref{section6} we propose two diagnosis as probe to
distinguish between the $\Lambda$CDM and the alternative models.
Section \ref{conc} concludes the paper.

\section{modified gravity in Palatini formalism}
\label{section2}

For $f(R)$ gravity, there are two main approaches to obtain the
field equation. The first one is so-called ''metric formalism''
which is obtained by the variation of the action with respect to
the metric. In this case the derived field equation is a fourth
order nonlinear differential equation. In the second approach,
which is called Palatini formalism, the connection and metric are
considered as independent fields and the variation of action with
respect to these fields results in a set of second order
differential equations. The Palatini formalism
%supports second
%order differential equations for the field, which will makes it a
is a plausible candidate to be the effective classical theory of
gravity from a more fundamental theory of Loop Quantum Gravity
\cite{sotiriou09}.

Let us take a general form of the action in Palatini formalism as:
\begin{equation}
S[f;g,\hat{\Gamma},\Psi_{m}]=-\frac{1}{2\kappa}\int{d^{4}x\sqrt{-g}f(R)+S_{m}[g_{\mu\nu},\Psi_{m}]},
\end{equation}
where $\kappa=8\pi{G}$ and $S_{m}[g_{\mu\nu},\Psi_{m}]$ is the
action of matter depends on the metric $g_{\mu\nu}$ and the
matter field $\Psi_{m}$.
$R=R(g,\hat{\Gamma})=g^{\mu\nu}{R_{\mu\nu}}(\hat{\Gamma})$ is the
generalized Ricci scalar and $R_{\mu\nu}$ is the Ricci tensor made
of affine connection. Varying  action with respect to the metric
results in:
\begin{equation}
f'(R)R_{\mu\nu}(\hat{\Gamma})-\frac{1}{2}f(R)g_{\mu\nu}=\kappa
T_{\mu\nu}, \label{field}
\end{equation}
where prime is the derivative with respect to the Ricci scalar and
$T_{\mu\nu}$ is the energy momentum tensor
\begin{equation}
T_{\mu\nu}=\frac{-2}{\sqrt{-g}}\frac{\delta S_{m}}{\delta
g^{\mu\nu}}.
\end{equation}
On the other hand varying the action with respect to the
connection results in:
\begin {equation} \label{connection}
\hat{\nabla_{\alpha}}[f'(R)\sqrt{-g}g^{\mu\nu}]=0,
\end {equation}
where $\hat{\nabla}$ is the covariant derivative defined from
parallel transformation and is given by the affine connection.
From equation (\ref{connection}), we define a new metric,
$h_{\mu\nu}=f'(R)g_{\mu\nu}$ conformally related to the physical
metric where the connection is the Christoffel symbol of this new
metric.

We apply flat FRW metric (namely $K=0$) for the universe
\begin{equation}
ds^{2}=-dt^{2}+a(t)^{2}\delta_{ij}dx^{i}dx^{j},
\end{equation}
and assume that universe is filled with a perfect fluid with the
energy-momentum tensor of $T^{\nu}_{\mu}=diag(-\rho,p,p,p)$. Using
the metric and energy momentum tensor in Eq.(\ref{field}), we
obtain the generalized FRW equations. It should be noted that the
conservation law of energy-momentum tensor,
$T^{\mu\nu}{}_{;\mu}=0$ is satisfied according to the covariant
derivative with respect to the metric and this definition
guarantees the motion of particles on geodesic \cite{koiv06}.
Combination of $G_{00}$ and $G^i_i$ results in\cite{Alemandi}:
\begin{equation}\label{Hub1}
(H+\frac{1}{2}\frac{\dot{f'}}{f'})^{2}=\frac{1}{6}\frac{\kappa(\rho+3p)}{f'}+\frac{1}{6}\frac{f}{f'}.
\label{hp}
\end{equation}
Taking the trace of equation (\ref{field}) results in
\begin{equation}
Rf'(R)-2f(R)=\kappa T, \label{trace}
\end{equation}
where $T=g^{\mu\nu}T_{\mu\nu}=-\rho+3p$. The time derivative of
this equation results $\dot{R}$ in terms of time derivative of
density and pressure. Using the equation of state of cosmic fluid
$p=p(\rho)$ and the continuity equation, the time derivative of
Ricci obtain as:
\begin{equation}
\dot{R}=3\kappa H\frac{(1-3dp/d\rho)(\rho+p)}{Rf''-f'(R)}.
\label{rdot}
\end{equation}
To obtain generalized first FRW equation we start with equation
(\ref{trace}) and obtain the density of matter in terms of Ricci
scalar as:
\begin{equation} \label{den}
\kappa\rho=\frac{2f-Rf'}{1-3\omega},
\end{equation}
where $w=p/\rho$. We substitute equation (\ref{den}) in (\ref{hp})
and use equation (\ref{rdot}) to change the time derivative to
${d}/{dR}$. We rewrite equation (\ref{hp}) as follows:
\begin{equation}
H^{2}=\frac{1}{6(1-3\omega)f'}\frac{3(1+\omega)f-(1+3\omega)Rf'}{\left[1+\frac{3}{2}(1+\omega)\frac{f''(2f-Rf')}{f'(Rf''-f')}\right]^{2}}.
\label{hpala}
\end{equation}
On the other hand using equation ({\ref{trace}) and continuity
equation, the scale factor can be obtained in terms of Ricci
scalar
\begin{equation}
a=\left[\frac{1}{\kappa\rho_{0}(1-3\omega)}(2f-Rf')\right]^{-\frac{1}{3(1+\omega)}},
\label{apala}
\end{equation}
where $\rho_{0}$ is the energy density and $a_{0}$ is the scale
factor (set to one, i.e.$~a_0 = 1$) at the present time. Now for a
generic modified action, omitting Ricci scalar in favor of the
scale factor between equations (\ref{hpala}) and (\ref{apala}) we
can obtain the dynamics of universe (i.e. $H=H(a)$).

For the simple case of matter dominant epoch $\omega=0$ which is
in our concern, these equations reduce to:
\begin{equation} \label{Hub}
H^{2}=\frac{1}{6f'}\frac{3f-Rf'}{\left[1+\frac{3}{2}\frac{f''(2f-Rf')}{f'(Rf''-f{'})}\right]^{2}}
\end{equation}
\begin{equation} \label{scalefactor}
a=\left[\frac{1}{\kappa\rho_{0}}(2f-R f')\right]^{-\frac{1}{3}}
\end{equation}

\section{Inverse Method in Palatini formalism}
\label{section3} In this section we introduce the inverse method
to extract $f(R)$ action in Palatini formalism from the dynamics
of the universe. This method has been studied in the work by
Rahvar and Sobuti in the metric formalism \cite{rahvar} and we
extend it to the Palatini formalism.
%and extract the
%appropriate action from the observational data.
%The idea is the
%possibility of extraction of the action through the dynamics of
%the universe, means knowing $a(t)$ or equivalently $H(z)$.

Replacing $f$ with the first derivatives of action from the
equation (\ref{trace}),
\begin{equation} \label{act}
f(R)=\frac{1}{2}[{RF}-\kappa(3p - \rho)],
\end{equation}
equation (\ref{Hub1}) can be written as follows:
\begin{equation} \label{equation15}
(H+\frac{1}{2}\frac{\dot{F}}{F})^{2}=\frac{R}{12} +
\frac{\kappa}{4F}(\rho+p).
\end{equation}
where $F$ is defined as $F={df}/{dR}$. It should be noted that
Ricci tensor in the Palatini formalism is given in terms of the
conformal metric of $h_{\mu\nu} = F(R) g_{\mu\nu}$. Substituting
the new metric in the definition of Ricci scalar results in:
\begin{equation} \label{Ricci}
R_{\mu\nu}=R_{\mu\nu}(g)+\frac{3}{2}\frac{\nabla_{\mu}F\nabla_{\nu}F}{F^{2}}-
\frac{\nabla_{\mu}\nabla_{\nu}F}{F}-\frac{1}{2}g_{\mu\nu}\frac{\nabla_{\alpha}\nabla^{\alpha}F}{F},
\label{riccit}
\end{equation}
where $R_{\mu\nu}(g)$ is the Ricci tensor defined in terms of
metric $g_{\mu\nu}$. By taking trace from equation (\ref{riccit})
we obtain relation between the Ricci scalar in Palatini and
metric as:
\begin{equation} \label{Ricpal}
R=R(g)+\frac{3}{2}\frac{\nabla_{\alpha}F\nabla^{\alpha}F}{F^{2}}-\frac{3\nabla_{\alpha}\nabla^{\alpha}F}{F}.
\end{equation}
On the other hand using the FRW metric, $R(g)$ as the Ricci scalar
in metric formalism is given by the Hubble parameter as:
\begin{equation}\label{ric}
R(g)=6\dot{H}+12H^{2},
\end{equation}
where for simplicity in the calculation we rewrite this equation
by changing the time derivative to the redshift derivative. In
what follows we use prime for the derivative with respect to the
redshift.
\begin{equation}\label{ric}
R(g)=-6HH'(1+z)+12H^{2}.
\end{equation}

For the Ricci scalar in Palatini formalism from equation
(\ref{Ricpal}), we obtain:
\begin{eqnarray} \label{Ricci1}
& & R
=-6HH'(1+z)+12H^{2}-\frac{3}{2}H^2(1+z)^2(\frac{F'}{F})^2\nonumber
\\ &+&3(1+z)^2HH'\frac{F'}{F}-6H^2(1+z)\frac{F'}{F}
+3H^2(1+z)^2\frac{F''}{F}
\end{eqnarray}

Substituting this equation in (\ref{equation15}) results in a
differential equation for the evolution of $F$ as a function of
redshift:
\begin{eqnarray}
 F'' &-& \frac32\frac{F'^2}{F} + F'(\frac{H'}{H} + \frac{2}{1+z})
\nonumber \\
&-&\frac{2H'}{H(1+z)}F + \frac{\kappa}{H^2(1+z)^2}(\rho+p) = 0.
\label{difftau}
\end{eqnarray}
In the matter dominant epoch we rewrite this equation by putting
$p = 0$ and $\rho = \rho_0(1+z)^3$, where we can replace
$\kappa\rho_0$ with $3H_0^2\Omega_{m}$. It should be noted that
the definition of the $\Omega_m$ is different from that in the
standard FRW equations.

For a given dynamics of universe (i.e. $a(t)$), we can extract the
Hubble parameter in terms of redshift, applying it in equation
({\ref{difftau}}) will provide $F$ in terms of redshift. On the
other side we can calculate the Ricci scalar from equation
({\ref{Ricci1}}) in terms of redshift. Eliminate $z$ in favor of
Ricci scalar results in $F(R)$. Finally by the numerical
integration of this function we can obtain the modified gravity
action, $f(R)$.

In the rest of this section to test this algorithm we use an
ansatz for the scale factor and try to extract the corresponding
action from a given dynamics. We apply the following ansatz for
the scale factor proposed in \cite{rahvar}:
\begin{equation}
a(\tau)=\frac{1}{1+p}(\tau)^{\frac{2}{3}}[1+p\tau^{\frac{2\alpha}{3}}]
\end{equation}
in which $\tau=tH_{0}$ is a dimensionless time parameter defined
in the interval of  0 to 1 and $H_{0}$ is the Hubble parameter at
the present time. This proposed dynamics has two free parameters
of $\alpha$ and $p$. We obtain the corresponding Hubble parameter
from this scale factor and consequently the distance modulus and
compare the model with the observed SNIa data. The best value for
the parameters of model has been obtained roughly,
$p=\frac{1}{3}$ and $\alpha=2$ in \cite{rahvar}. Substituting the
scale factor in equation (\ref{difftau}), we obtain the deviation
parameter of $\xi \equiv F - 1$ from GR as a function of
redshift, which is plotted in Figure ({\ref{fig1}).
\begin{figure}[t] \label{fig1}
\begin{center}
\epsfxsize=9.0truecm \epsfbox{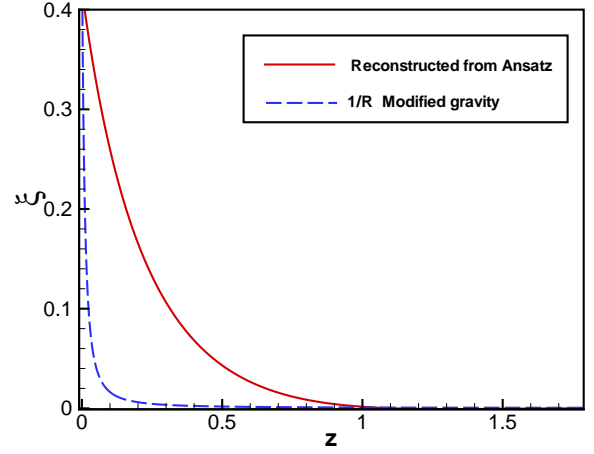} \narrowtext \
\caption{Deviation parameter $\xi=F(R)-1$ is obtained from the
ansatz scale factor (solid-line) is compared with $\xi$-parameter
of $1/R$ modified gravity (dashed-line) in terms of redshift. For
the redshifts $z>1$, the $f(R)$ gravity model converge to the
Einstein-Hilbert action.} \label{fig1}
\end{center}
\end{figure}
\begin{figure}[t] \label{fig2}
\begin{center}
\epsfxsize=9.0truecm \epsfbox{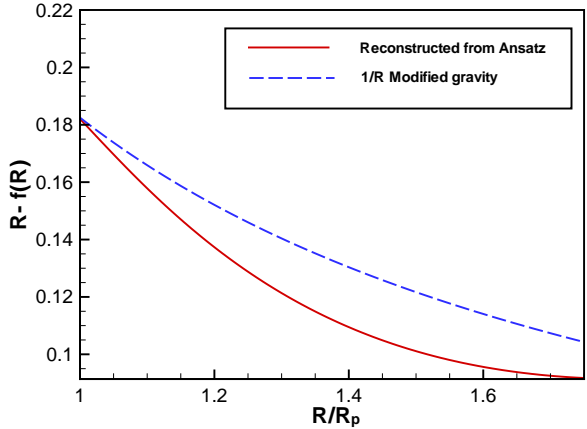} \narrowtext \
\caption{The difference of modified gravity action from the
Einstein-Hilbert action is plotted versus Ricci scalar
(solid-line) and compared with the action of
$f(R)=R-\frac{\mu^4}{R}$ (dashed-line). The Ricci scalar is
normalized to its value at the present time, $R_p$.} \label{fig2}
\end{center}
\end{figure}
In order to reconstruct the action in terms of Ricci action, we
obtain the Ricci scalar in terms of redshift and finally by
eliminating redshift between $F$ and Ricci scalar, we obtain
$F=F(R)$. Integrating this function provides action in terms of
Ricci scalar. We also plot the $\xi$-parameter of $1/R$ modified
gravity for comparison with the reconstructed modified gravity in
Figure (\ref{fig1}) and $R-f(R)$ as a function of $R$ in Figure
(\ref{fig2}). Comparison of the extra term with the
Einstein-Hilbert action roughly resembles to $\mu^{4}/R$ function
with $\mu^{2}\sim 10^{-1}R_{p}$ where $R_{p}$ is the present
value of Ricci scalar.

% So
%we can  approximately express the action of this dynamics with
%well known function of
%\begin{equation}
%f(R)=R-\frac{\mu^{4}}{R},
%\end{equation}
%where $\mu^{2}\sim 10^{-1}R_{p}$ where $R_{p}$ is the present value
%of Ricci scalar.
%

\section{Reconstruction of the dynamic of the Universe by SNIa data}
\label{section4} In section \ref{section3}, we showed that it is
possible to reconstruct the  modified gravity action by knowing
the dynamics of the universe. In this section, we use SNIa
cosmological data to obtain the Hubble parameter and consequently
reconstruct the action of modified gravity.

The dynamics of Hubble parameter, $H(z)$, can be obtained if the
distance modulus of Supernovas data as a function of redshift is
known. In FRW--flat universe, the Hubble parameter is related to
the distance modulus of SNIa as follows:
\begin{equation}
H(z)=[\frac{d}{dz}(\frac{d_{L}(z)}{1+z})]^{-1}.
\label{rec}
\end{equation}
The main challenging point in this procedure is the limited number
of observed SNIa which impose an uncertainty in calculating the
continues function for $H(z)$. The overall number of Supernovas
which has been detected is in the order of 300-400. We use the
latest Supernova data of the Union sample to extract the Hubble
parameter \cite{Kowalski}. To make a continues Hubble parameter,
we follow the procedure known as reconstruction method, proposed
by Shafieloo et al. in \cite{shafieloo}}. In this algorithm, a
non-parametric function is used for smoothing the distance
modulus of supernova data over the redshift. Here we choose a
guess model resemble to the observed distance modulus of the
supernova data. In the next step we subtract the distance modulus
of the observed data from the guess model using a gaussian
function for smoothing the observed data as follows:
\begin{equation}
\mu^{d}(z)=N(z)\sum_i[\mu^{g}(z)-\mu^{obs}(z_{i})]\exp{\frac{-(z-z_{i})^{2}}{2\Delta^{2}}},
\label{smooth1}
\end{equation}
where
\begin{equation}
N(z)^{-1}=\sum_i \exp[\frac{-(z-z_{i})^2}{2\Delta^2}],
\end{equation}
where $z_{i}$ represents the redshift of each SNIa in the Union
sample. The sum term is considered for all 307 Union sample data,
$\mu^{obs}(z_{i})$ is the observed distance modulus and
$\mu^{g}(z)$ is a continues guess model for the distance modulus.
$N(z)$ is a normalization factor and $\Delta$ is a suitable
redshift window function. $\mu^{d}(z)$ is a smoothed continues
function for the residual of luminosity distance in terms of
redshift. Now the corrected distance modulus is added to the
guess model to generate the new smooth function for the distance
modulus:
\begin{equation} \label{smooth}
\mu^{s}(z)=\mu^{g}(z)+\mu^{d}(z).
\end{equation}
We repeat this procedure using $\mu^s(z)$ as the new guess
function. It can be shown that after a finite time of this
irritation, $\chi^2$ of the smoothed function with respect to
observed data will converge to a fixed value. It means that we
will find a continues distance modulus function with the best fit
to the real data. It is shown in \cite{shafieloo} that the result
of the best continues function is independent of the choice of
the first guess model. Having the the smoothed luminosity
distance we use equation (\ref{rec}) to obtain the Hubble
parameter H(z). Another point is that the results will clearly
depend upon the value of $\Delta$ in equation (\ref{smooth1}). A
large value of $\Delta$ produce a smooth result, but the accuracy
of reconstruction worsens, while small value $\Delta$ gives more
accurate, but noisy result. Considering the frequency of SNIa
data observed in the Union sample, we choose
$\Delta=\sqrt{M}(1+z)\Delta_{0}$ for the window function, where M
is the number of irritation need to converge $\chi^2$ and
$\Delta_{0}=N^{-1/3}$ where N is the number of SNIa
\cite{shafieloo}.

\begin{figure}[t]
\epsfxsize=9.0truecm
\begin{center}
\epsfbox{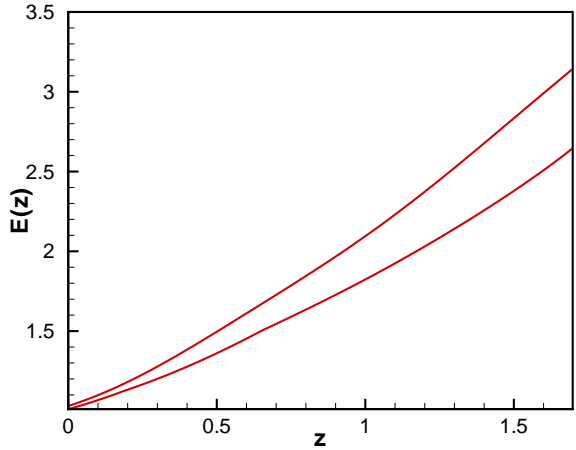} \narrowtext \caption{ The uncertainty in
$E(z)=\frac{H}{H_{0}}$ from the smoothing method, resulting from
the Monte-Carlo simulation. We generate one hundred realization
for the distance modulus and obtain the Hubble parameter which all
these continues functions reside inside the boundaries in the
figure.} \label{fig3}
\end{center}
\end{figure}
In what follows we find the uncertainty in $H(z)$ from this method
to use it for reconstructing the appropriate action in the
Palatini formalism. For this purpose we do a Monte-Carlo
simulation, generating 100 realization of SNIa data and using the
same distribution of SNIa in terms of redshift reported by
Kowalski et al.\cite{Kowalski}. Also we use the same error bars of
distance modulus in the observed data. In order to simulate the
synthetic distance modulus of SNIa, we assume a dark energy model
for the universe with a constant equation of state of
$\omega=-0.75$. Choosing $\Lambda$CDM as the guess model for these
data, we obtain the Hubble parameter $E(z) = H(z)/H_0$ for $100$
realization of supernova data. Figure (\ref{fig3}) shows the
boundaries for $E(z)$, resulted from 100 realization of Supernova
data.

Now we can apply this uncertainty to $H(z)$, resulting from the
smoothing procedure on the real Union sample \cite{Tegmark}. We
plot the Hubble parameter from the Union sample as shown in Figure
(\ref{fig4}) with a margin represents the uncertainty that is
obtained form the Monte-Carlo simulation. Using the method
described in section \ref{section3}, we extract the corresponding
modified gravity in Palatini formalism. The parameter of $\xi =
{df(R)}/{dR}-1$ in terms of Ricci scaler shows a small deviation
from the Einstein-Hilbert action as shown in Figure (\ref{fig5})
with the uncertainty of this parameter. Here the deviation from
the Einstein-Hilbert action is in the order of $\xi\sim 10^{-3}$.
We ask the reliability of this result in terms of the number of
SNIa data. In the next section we will discus about this issue.

\begin{figure}[t]
\begin{center}
\epsfxsize=9.0truecm \epsfbox{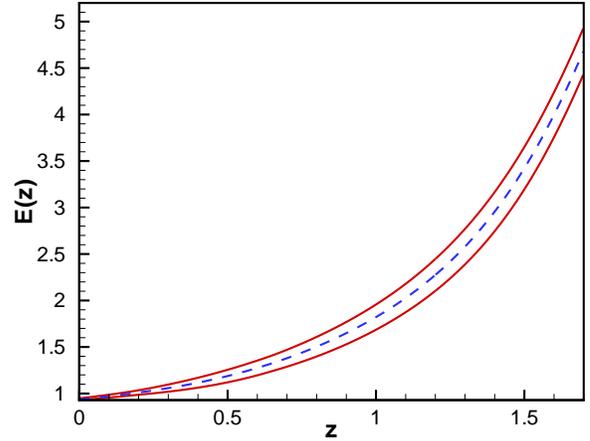} \narrowtext \ \caption{
Dashed line is the reconstructed Hubble parameter from SNIa Union
sample data. Bold lines are the confidence level of Hubble
parameter.}\label{fig4}
\end{center}
\end{figure}

\begin{figure}[t]
\begin{center}
\epsfxsize=9.0truecm \epsfbox{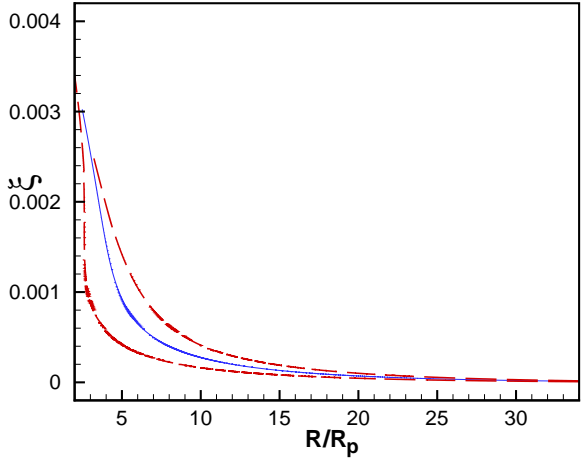} \narrowtext \ \caption{ The
bold line shows the reconstructed modified gravity $\xi\equiv
F-1$ parameter versus Ricci scalar normalized to its present
value.  The dashed lines show the confidence level of the $\xi$
parameter corresponds to the uncertainty of the Hubble parameter
resulting from the smoothing procedure.}\label{fig5}
\end{center}
\end{figure}

\section{Viability of smoothing method and the number of SNIa data}
\label{section5} In this section we examine the viability of
smoothing of the Hubble parameter. The results will be applied to
the modified gravity models in the next section.

According to the methodology of the smoothing procedure, the
Hubble parameter depends on the quantity and quality of SNIa
data. Similar to the simulation in the previous section we
generate 100 realization of 307 SNIa data, using the redshift and
the uncertainty of the distance modulus in the Union sample
within the framework of $\Lambda$CDM model. We generate continues
Hubble parameter with the margin of uncertainty which is shown in
Figure (\ref{fig6}).
% represents the domain that the Hubble parameter from 100 realization reside in it.
On the other hand we obtain the Hubble parameter in $\Lambda$CDM
model though fitting the observed distance modulus of the data to
the model as shown in Figure (\ref{fig6}). The margin represents
1$\sigma$ level of confidence for the Hubble parameter. Comparing
the uncertainties from these two methods show that the smoothing
method is not reliable with 307 number of supernova data.

We increase the number of SNIa data with the quality of Union
sample to have the same uncertainty in the Hubble parameter from
the two methods. For $N>1500$ as shown in Figure (\ref{fig6}) we
can rely on the smoothing method. We can repeat this simulation
with precise data than the Union sample to achieve this goal with
a smaller number of Supernova data.

\begin{figure}[t]
\begin{center}
\epsfxsize=9.0truecm \epsfbox{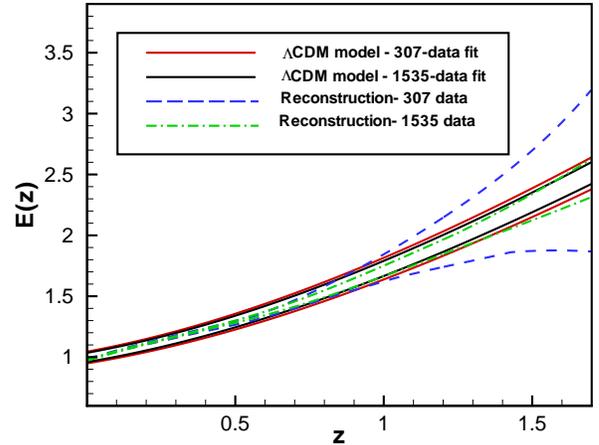}
\narrowtext \ \caption{ Dashed lines and dotted dashed line are
the reconstructed Hubble parameters from 307 and 1535 SNIa data,
respectively. Solid lines are 1-$\sigma$ confidence level of
Hubble parameter from data fitting to 307 and 1535 data in
$\Lambda$CDM model.}\label{fig6}
\end{center}
\end{figure}

\section{Distinguishing between the Modified Gravity models and $\Lambda$CDM}
\label{section6}

In this section we present two diagnosis in order to distinguish
between $\Lambda$CDM from an alternative model. The first method
is (a) Comparison of the Hubble parameters and the second method
is (b) the $\Xi$-function. Both of diagnosis are applicable by
using the inverse method.

\subsection{Comparison of the Hubble parameters}
In the previous section we have seen the smoothing method to
generate a continues Hubble parameter from the observed data. If
the dynamics of universe follows rather than the $\Lambda$CDM
model, can we distinguish the real model of the universe, knowing
the Hubble parameter from the data ?

We can compare the reconstructed Hubble parameter directly from
the observed data with that of obtained from fitting data to the
$\Lambda$CDM model. We subtract the two Hubble parameters and
compare it with $1\sigma$ deviation of the smoothed Hubble
parameter. As we simulated in the previous section using $N>1500$
is sufficient to reliable on the uncertainty of the Hubble
parameter.

We use $f(R)=R-\frac{\mu^4}{R}$ action in Palatini formalism and
generate $1535$ supernova data with the quality of the Union
sample. Using the smoothing method we obtain the Hubble parameter.
On the other hand we fit the generated data with the $\Lambda$CDM
model. Figure (\ref{fig8}) compares the two Hubble parameters and
shows that for the redshifts with $z>1$, the two models are
distinguishable.
\begin{figure}[t]
\begin{center}
\epsfxsize=9.0truecm \epsfbox{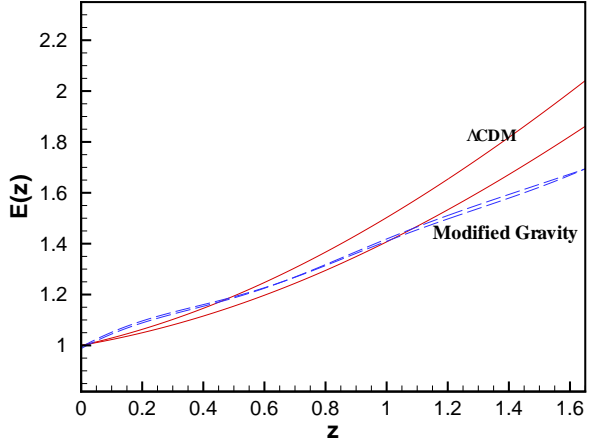} \narrowtext \
\caption{\label{fig8} The solid-line is the Hubble parameter with
1-$\sigma$ error bar from fitting the simulated data to the
$\Lambda$CDM model. The dashed-line is the confidence level of
Hubble parameter extracted from the smoothing method of the
Hubble parameter. The figure shows the deviation of the two curve
for $z>1$.} \label{age}
\end{center}
\end{figure}

\subsection{The method of $\Xi$-function}
We define a new criterion to distinguish the $\Lambda$CDM from an
alternative model as follows:
\begin{equation}
\Xi\equiv\frac{2E(z)E^{'}(z)}{3(1+z)^2},
\end{equation}
where prime is differential with respect to the redshift. For the
case of $F=1$ in $\Lambda$CDM universe, from equation
(\ref{difftau}), $\Xi$ reduces to $\Xi = \Omega_{m}^{0}$. If we
have the smoothed normalized Hubble parameter from the observed
data, $E(z)$, then $\Xi$ can be obtained. Any deviation from a
constant value for this function is a diagnosis for the
$\Lambda$CDM universe. This method is applicable both for the
dark energy and modified gravity models. Let us take $\omega(z)$
as the equation of state of dark energy, we can use FRW equations
to obtain $\Xi$ as follows:
\begin{equation}
\Xi=\Omega_{m}^{0}+(1-\Omega_{m}^{0})\frac{1+\omega(z)}{(1+z)^3},
\end{equation}
where for the constant value of the equation of state,
$\omega(z)=-1$, this equation reduces to $\Xi=\Omega_{m}^{0}$.

%Also we can identity either the equation of state is a
%quintessence model $-1<\omega(z)<-1/3$ or that is a phantom model
%$\omega<-1$.

In order to show the deviation from the Einstein-Hilbert action,
we can calculate $\Xi-\Omega_m^{(0)}$ as a function of redshift.
Here we know $\Omega_m^{(0)}$ from a direct cosmological
observations such as gravitational weak lensing.
\begin{figure}[t]
\begin{center}
\epsfxsize=9.0truecm \epsfbox{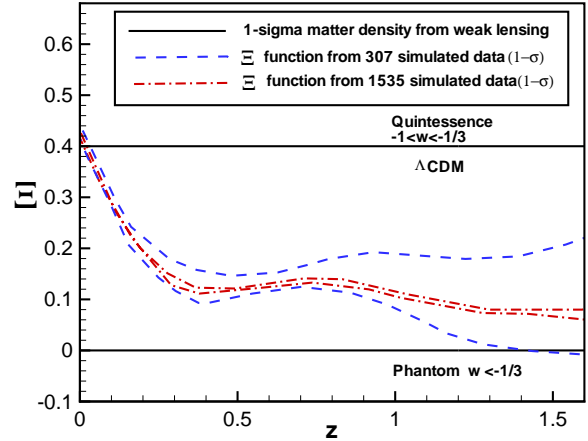} \narrowtext \
\caption{\label{fig9} $\Xi$-function versus redshift is plotted
for two series of SNIa data, with 1$\sigma$ level of confidence.
The dash-lines and dashed-dotted lines indicate for 307 and 1535
data respectively. Solid lines indicate the matter density of
universe derived from gravitational weak lensing.}
\end{center}
\end{figure}
To show how this method works we generate SNIa data from $f(R) =
R - \mu^4/R$ action resemble to the Union sample and after
smoothing the Hubble parameter, we plot $\Xi$ as a function of
redshift. We generate $\Xi$ function for two cases with $307$ and
$1535$ supernova data as represented in Figure (\ref{fig9}). To
compare this function with a direct measurement of
$\Omega_m^{(0)}$, we use the weak lensing data which puts limit
on the matter content of the universe in the range of
$\Omega_m^{(0)}=0.2 \pm^{0.2}_{0.2}$\cite{Waer2001}. As shown in
Figure (\ref{fig9}), increasing of SNIa data will pin down the
$\Xi$ function more precisely where today's SNIa data is not
sufficient. On the other hand direct observations of matter
density of universe are not much precise to claim any model
distinguishing results.

\section{conclusion}
\label{conc}

One of the most puzzling questions in the cosmology is the
physical mechanism for the acceleration of the universe. Is it
driven by a cosmological constant or universe is filled with an
exotic dark energy or the Einstein gravity should be modified? In
this work we proposed an inverse method to extract the action of
a modified gravity in the Palatini formalism from the expansion
history of universe.

We used the smoothing method to obtain a continues Hubble
parameter from the Supernova Type Ia Union sample data. We showed
that more than 1500 supernova data with the quality of union
sample is essential to have a reliable Hubble parameter. Finally
we proposed two cosmological diagnosis in order to distinguish
between $\Lambda$CDM and alternative models. The first one
compares the smoothed Hubble parameter from the SNIa data with the
standard $\Lambda$CDM model. In the second approach we define a
new parameter where it is constant, equal to the $\Omega_m$ for
the $\Lambda$CDM universe and vary with the redshift for any
alternative model. A precise measurement of matter content of the
universe from one hand an enough number of Supernova data from
the other hand will enable us to identify either our universe
follows $\Lambda$CDM universe or some modification is necessary.

%********************************************

\end{document}